\documentstyle[12pt]{article}
\font \msb=msbm10 scaled \magstep1
\newcommand{\bR}{\mbox{\msb R} }
\def\l{{\lambda}}
\def\a{{\alpha}}
\def\b{{\beta}}
\def\g{{\gamma}}
\def\d{{\delta}}
\def\D{{\Delta}}
\def\x{{\xi}}
\def\SO{{\cal O}}
\def\te#1{{\widetilde{#1}}}

\def\nn{ \nonumber }
\def\bq{ \begin{equation} }
\def\eq{ \end{equation} }
\def\ben{ \begin{eqnarray} }
\def\en{ \end{eqnarray} }
\def\ll{ \label }
\def\frac#1#2{{#1\over #2}}
\def\dfrac#1#2{{\displaystyle{#1\over#2}}}
\def\dpr#1#2{{\displaystyle{{\partial #1}\over{\partial#2}}}}

\newtheorem{prop}{Proposition}
\begin{document}
\title{Quasi-point separation of variables for
the Henon-Heiles system and a system with quartic potential.}
\author{
S. Rauch-Wojciechowski\\
{\small\it Department of
Mathematics,  Linkoping University,
S-58183,  Linkoping,  Sweden}\\
 A.V. Tsiganov\\
{\small\it
 Department of Mathematical and Computational Physics,
 Institute of Physics,}\\
{\small\it
St.Petersburg University,
198 904,  St.Petersburg,  Russia}
}
\date{}
\maketitle

\begin{abstract}
We examine the problem of integrability of
two-dimensional Hamiltonian systems by means of
separation of variables.
The systematic approach to construction of
the special non-pure coordinate separation of variables for
certain natural two-dimensional Hamiltonians is proposed.
\end{abstract}


\section{Introduction}
In this note we study quasi-point separation of
variables for certain natural Hamiltonians
\bq
H=\dfrac12(p_x^2+p_y^2)+V(x,y)
\ll{nh}
\eq
of two degrees of freedom.

The classical separability theory \cite{w70,ll60} is
concerned with orthogonal point transformations $x=\Phi_1(u,v)$
and $y=\Phi_2(u,v)$ for which the corresponding Hamilton-Jacobi
equation expressed in terms of $(u,v)$-variables can be solved
by separation of variables.  The list of coordinate systems in
plane that provide point separation of variables for
systems with the Hamiltonian (\ref{nh}) are well known.  Even an
effective criterion of separability for a given potential
$V(x,y)$ has been formulated \cite{mw88,w70}.  In this note we
consider a non-pure point transformation related to a cartesian
system of coordinates.

Very little is known about general canonical transformations
$x_k=\Phi_k(u_j,p_{u_j})$ and $p_k=\Psi_k(u_j,p_{u_j})$,
$k=1,2$, which separate the Hamiltonian (\ref{nh}).  Recently,
a special type of non-point transformations
$x=\varphi(H,C)\cdot\Phi(u,v)$ has been introduced for the
third integrable case of the Henon-Heiles system
\cite{rav93,brw94} and for a system with a quartic potential
\cite{rav94}. Here $\varphi(H,C)$ is a certain function of the
Hamiltonian $H$ and of the second integral of motion $C$.  On
the orbit $\SO$ ($H=\a_1\,\quad C=\a_2$) this transformation
becomes a point transformation and, therefore, we shall call it
a quasi-point transformation.  These transformations have been
found either in the context of the Painlev\'e expansion
\cite{rav93,rav94} or have been derived from the Miura
transformation for the systems related to stationary flows of
soliton equations \cite{brw94,bef95}.  However no general
principle of constructing general (this means non point)
canonically separable potentials is known.  This is the reason
why in this note we address to this question directly by
starting with general separated equations and by trying to find
transformations which lead to natural Hamiltonians (\ref{nh})
of two degrees of freedom.  We do not find any essentially new
potentials.  We present in detail our construction of
separating variables and, then, list the main results.  We
explain the connection of quasi-point transformations with the
supersymmetric quantum mechanics.


\section{Quasi-point canonical transformation}
\setcounter{equation}{0}

Let us begin with a two-dimensional system expressed in terms of
canonical variables of separation $(u,p_u,\,v,p_v)$  with the
rectangular separated equations of the general form
\cite{ll60,w70}
\ben
&\D_1(u,p_u)&=f(u)p_u^2+V_1(u)\,,\ll{sepeq}\\
&\D_2(v,p_v)&=g(v)p_v^2+V_2(v)\,.\nn
\en
The functions $\D_j$ Poisson commute $\{\D_1,\D_2\}=0$ with
respect to the standard Poisson brackets.  Notice that we
could, by the using a point  transformation, reduce these
equations to equations with $f(u)=g(v)=1$, but the form
(\ref{sepeq}) is more suitable for further calculations.

As commuting integrals of motion we choose
\ben
H&=&\D_1+\D_2=
f(u)p_u^2+g(v)p_v^2+V_+(u,v)\,,\nn\\
&&\ll{int}\\
C&=&a\,\D_1-b\,\D_2=
a\, f(u)p_u^2-b\, g(v)p_v^2+V_-(u,v)\,,\qquad a,b\in{\bR}\nn
\en
which clearly define  $V_\pm(u,v)$.  These integrals are second
order polynomials in momenta.  A condition of separability for
the potentials $V_\pm$ is
\bq
\dfrac{\partial^2}{\,\partial u\,\partial v\,}\, V_\pm(u,v)=0\,.
\ll{ddv}
\eq

Let as denote a canonical transformation
to the cartesian variables $(x,y)$ as
\bq
x=\Phi_1(u,p_u,v,p_v)\qquad{\rm and}\qquad
y=\Phi_2(u,p_u,v,p_v)\,,\ll{cantr}
\eq
and let us require that after the transformation (\ref{cantr})
the Hamiltonian
(\ref{int}) takes the natural Hamiltonian form (\ref{nh}).
Then
\[p_x=\{H,x\}\qquad {\rm and}\qquad p_y=\{H,y\}\]
and, after substituting these momenta into the canonical Poisson
brackets, we shall get a system of equations for the
transformations (\ref{cantr}).  It reads
\ben
\{\Phi_j(u,p_u,v,p_v),\Phi_k(u,p_u,v,p_v)\}&=&0\,,\nn\\
\{\{H,\Phi_j(u,p_u,v,p_v)\},\{H,\Phi_k(u,p_u,v,p_v)\}\}&=&0\,,\ll{restr}\\
\{\{H,\Phi_j(u,p_u,v,p_v)\},\Phi_k(u,p_u,v,p_v)\}&=&\d_{jk}\,,\nn\\
 (j,k=1,2)\,.\qquad&&\nn
\en
Since, we can not solve this system of equations in the whole
generality, we have to make certain simplifying ansatz for the
functions $\Phi_k$ in order to obtain certain particular solutions of
(\ref{restr}).

In the class of  point transformations
$\te{x}=\te{\Phi}_1(u,v)$ and $\te{y}=\te{\Phi}_2(u,v)$ the
general solution of (\ref{restr}) has the form
\bq
\te{x}=\a u+\b v+ \l_1\,,\qquad \te{y}=\g u+ \d v+ \l_2\,,
\ll{rot}
\eq
which is a superposition of rotation and  translation
transformations.  Since the functions $f(u)$, $g(v)$ and
$V_\pm(u,v)$ are arbitrary we can always rewrite the solution
(\ref{rot}) in the following nonsymmetric form
\bq
\te{x}=\te{\Phi}_1(u,v)=u-v\,,\qquad \te{y}=\te{\Phi}_2(u,v)=u+v+2\g\,,
\ll{rot1}
\eq
which will be used below.  The polynomial order of integrals of
motion (\ref{int}) remains unchanged under (\ref{rot1}).

In order to obtain a non-point transformation as a solution of
(\ref{restr}) we shall introduce a particular ansatz for
$x=\Phi_1(u,p_u,v,p_v)$, which is suggested by the results of
\cite{rav93,rav94}.  For explaining the structure of this
ansatz we begin with a simple example, which will be present
in detail below. Let us take
\ben
&&\D_j(\x)=\x^np_\x^2+V_j(\x)\,,\qquad\x=u,v\,,\nn\\
\ll{exx}\\
&&x^2=\D_1(u)u^{-m}\,,\qquad u^m=\D_1 x^{-2}\,,\nn
\en
and require that the Hamiltonian (\ref{int}) takes the natural
form (\ref{nh}), when expressed through the new variables
$(x,p_x)$. Then, we obtain
\[\dfrac{p_x}x=-mu^{n-1}p_u\]
and by applying the Poisson brackets (\ref{restr}) we get
\bq
\left\{\dfrac{p_x}x,x^2\right\}=2=
\left\{-mu^{n-1}p_u,\,u^{n-m}p_u^2+u^{-m}V_1(u)\right\}\,.\ll{eeq}
\eq
The kinetic part of this equation leads to a restriction for
the kinetic part of the Hamiltonian: $n=2-m$. The
potential part of (\ref{eeq}) gives
\[
u\dpr{V_1(u)}u-mV_1(u)=-2m^{-1}u^{2m}\,,\]
which has the solution
\bq
V_1=-2m^{-2}u^{2m}+\a  u^m=-2m^{-2}\D_1^2x^{-4}+\a\D_1x^2\,,
\qquad\a\in\bR\,.\nn
\eq
According to the definitions (\ref{nh}) and (\ref{exx})
the momenta $p_u$ is
\[p_u=-\dfrac{\D_1p_xx^{-3}}{mu}\,,\]
and, after substituting $u,p_u$ and $V_1(u)$, the separated
equation $\D_1$ (\ref{exx}) becomes
\[\D_1(x,p_x)= \dfrac{p_x^2}2-\dfrac{m^2}2
(x^4-\a x^2)\,,\qquad\a\in\bR\,.
\]

The second pair of variables $(y,p_y)$ follows from the
second separated equation $\D_2(v)$ and the Hamiltonian
(\ref{int}) transforms to the new Hamiltonian
\[
H=\dfrac12(p_x^2+p_y^2)+\a_1x^4+\a_2y^4+\a_3x^2+\a_4y^2 \,,
\qquad \a_k\in\bR\,,
\]
under non-point canonical transformation (\ref{exx}).

Next we consider a symmetric form of this transformation
related to the symmetric form of the second integral of motion
(\ref{int}).
Let us take
\bq
P_n(x)=\sum_{k=0}^n a_k x^k=C\phi(u-v)\,,\qquad a_k\in\bR\,,
\ll{ansx}\eq
where $P_n(x)$ is a $n$-th order polynomial, $C$ is an
integral of motion (\ref{int}) and $\phi(u-v)$ is a function of one
variable $z=u-v$.  On the orbit $\SO$  ($C=c=const$) the
transformation (\ref{ansx}) becames a point transformation and,
therefore, we shall call this transformation a quasi-point
transformation.

After substituting the ansatz (\ref{ansx}) into the
equation
\bq
\{\{H(u,p_u,v,p_v),\Phi_1(u,p_u,v,p_v)\},\Phi_1(u,p_u,v,p_v)\}=1
\ll{xpx}
\eq
we consider  terms at independent powers of momenta.  It
appears that a solution of the corresponding equation exist
only if $P_n(x)=x^2$ (up to a point transformation
(\ref{rot})) and we get one equation for the function $\phi(z)$
\bq
2\,\dfrac{d\phi(z)}{dz}=\phi(z)\,\dfrac{d^2\phi(z)}{dz^2}\,,
\qquad z=u-v\,,\ll{phi}
\eq
and a system of equations for
the functions $f(u)$ and $g(v)$
\ben
f(u)-bg(v)-b(u-v)\dfrac{dg(v)}{dv}&=&0\,,\nn\\ \ll{fgeq}\\
-af(u)+bg(v)+a(u-v)\dfrac{df(u)}{du}&=&0\,.\nn
\en
The general solutions for (\ref{phi} and \ref{fgeq}) are
\ben
&&\phi(z)=z^{-1}=\te{\Phi}_1^{-1}(u,v)\,,\ll{alph}\\
&&a=b=\a_1\,,\qquad f(u)=\a_2u+\a_3\,,\quad g(v)=\a_2v+\a_3\,,
\quad \a_k\in{\bR}\,.\nn
\en
Further, we obtain that
\[x^2=C\te{\Phi}_1^{-1}(u,v)\,,\]
where $\te{\Phi}_1^{-1}(u,v)$ is a point transformation
(\ref{rot1}) and, therefore, we can consider (\ref{ansx}) as a
natural generalization of a pure point transformation, which
can be applied to the other three types of coordinate separated
equations in the plain. For instance for the parabolic system
of coordinates we have to substitute
\[C=\dfrac{v^2\D_1-u^2\D_2}{u^2+v^2}\,,
\qquad \te{\Phi}_1(u,v)=u^2-v^2\,,\]
into the ansatz (\ref{ansx}).

From the remaining terms of (\ref{xpx}) (by taking into account
(\ref{phi}) and (\ref{fgeq})) we obtain that the potential
$V_-(u,v)$ obeys the following equation
\bq
\,\left((\a_2u+\a_3)\dfrac{\partial V_-}{\partial u}
-(\a_2v+\a_3)\dfrac{\partial V_-}{\partial v}\right)
-\dfrac{(\a_2(u+v)+2\a_3)}{\a_1(u-v)}V_-
=-\dfrac{2(u-v)^3}{\a_1(u-v)}\,.\ll{eqvm}
\eq
For a  separable potential
$\partial^2 V_-(u,v)/\partial u\partial v=0$
this equation has a
unique solution
\bq
V_{1,2}(\x)= \dfrac{\-2/\a_1 \x^3+\b_2 \x^2+\b_1 \x
+\b_0}{\a_2\x+\a_3}\,; \quad \x=u,v.
\ll{v12}
\eq
where $\b_k$ are arbitrary constants.

Next we have to determine the second variable
$y=\Phi_2(u,p_u,v,p_v)$ by using equations (\ref{restr}).
Again we are not able to find a general solution for $\Phi_2$
and we shall use the particular ansatz of \cite{rav93,rav94}
\bq
Q_n(y)=\sum_{k=0}^n b_ky^k=\psi_1(u+v)+\psi_2(x,p_x)\,,\qquad
p_y=\{H,y\}\,,\ll{ansy}
\eq
where $Q_n(y)$ is a polynomial of the order $n$ and $\psi_k$ are
unspecified as yet functions.

After substituting (\ref{ansy}) into the equations
\[\{x,Q_n(y)\}=0 \qquad{\rm and}\qquad
\{p_x,Q_n(y)\}=0\]
(recall that $p_x=\{H,x\}$)
we obtain
\bq
Q_n(y)=\g_1\left[\,u+v+\g_2-{\dfrac{\a_1}2}\,
\left({\dfrac{p_x}{x}}\right)^2
+{\dfrac{\a_2x^2}{2}}\right]\,.\ll{ansy2}
\eq
This ansatz is a superposition of a point transformation
$\te{\Phi}_2(u,v)$ (\ref{rot1}) with a term depending on the
first variables $(x,p_x)$.

By using the remaining equation $\{p_y,y\}=1$ we obtain  (up to
a point transformations (\ref{rot})) that
\bq
Q_n(y)=y^2 \qquad \g_1=-\dfrac{1}{\a_2}\,, \ll{polin}
\eq
and yet another differential equation of the second order for the
potential $V_+(u,v)$. We present here a partial form of this
equation with $\a_2=0$
\[\dfrac{\partial^2 V_+}{\partial z^2}=z\dpr{V_+}{z}\,,\qquad
z=u-v\,.
\]

This equation for $V_+$ is equivalent to the equation for $V_-$
(\ref{eqvm}) provided that the constants $\a_1$, $\a_2$, $\b_2$ and
$\g_2$ satisfy
\bq
\g_2+\a_1\b_2+2\a_3=0\,.
\ll{rest}
 \eq
Let us summarize these considerations as
\begin{prop}
A quasi-point transformation of the form
\ben
P_n(x)&=&\sum_{k=1}^n a_kx^k=\dfrac{C}{u-v}\,,\nn\\
Q_m(y)&=&\sum_{k=1}^m b_ky^k= \psi_1(u+v)+\psi_2(x,p_x)\qquad
a_k,b_k\in\bR\nn
\en
extended to a canonical transformation, transforms the Hamiltonian
\[H=f(u)p_u^2+g(v)p_v^2+V_1(u)+V_2(v)\]
into the natural Hamiltonian form
\[H=\dfrac12(p_x^2+p_y^2)+V(x,y)\,,\]
if and only if
\ben
x^2&=&\dfrac{C}{u-v}\,,\nn\\
\nn\\
\dfrac{p_x}{x}&=&-\dfrac{(\a_2u+\a_3)p_u-(\a_2v+\a_3)p_v}{u-v}
\ll{ptr}\\
\nn\\
y^2:&=&-\dfrac{1}{\a_2}
\left(u+v-\a_1\b_2-2a_3
-\dfrac{\a_1}2\left(\dfrac{p_x}{x}\right)^2
+\dfrac{\a_2x^2}2
\right)\,,\nn\\
\nn\\
yp_y&=&\dfrac{1}{\a_2}\left(
(\a_2u+\a_3)p_u-(\a_2v+\a_3)p_v\right.\nn\\
&+&\dfrac{p_x}{2x}
\left(\dfrac{\a_1p_x^2}{x^2}
+\left.\dfrac{6\a_3}{\a_2}-6\a_2y^2+\a_1\b_2-\a_2x^2\right)
\right)\,.\nn
\en
and the potentials
\[
V_{k}(\x)=
\dfrac{\-2/\a_1 \x^3+\b_2 \x^2+\b_1 \x +\b_0}{\a_2\x+\a_3}\,;
\quad k=1,~\x=u\quad {\rm or}\quad k=2,~\x=v.
\]
Constants $\a_j\,,\b_j\,,j=1,2,3$ are a six free parameters
of this transformation.
\end{prop}
An inverse transformation to (\ref{ptr}) reads
\ben
\x&=&\pm \dfrac{C}{2x^2}
+\dfrac{\a_1p_x^2}{4x^2}
-\dfrac{\a_2}4 (2y^2+x^2)-\g_2\,;\quad \x=u,v\,,\nn\\
\ll{invtr}\\
p_\x&=&\dfrac1{2(\a_2\x+\a_3)}
\left( \dfrac{p_xC}{x^3}-\a_2yp_y -\right.\nn\\
\nn\\
&-&\left.\dfrac{p_x}{2x}\left(
\dfrac{\a_1p_x^2}{x^2}+
\dfrac{6\a_3}{\a_2}-6\a_2y^2+\a_1\b_2-\a_2x^2\right)\right)\,,
\nn
\en
where the integral $C$ is an unspecified as yet function of
the new variables $(x,p_x,y,p_y)$.  We have to substitute this
inverse transformation into the definition of the integral of
 motion $C$ (\ref{int}) and to solve the resulting equation.
It has the polynomial form
\bq
aC^4+bC^2+c=0\,,\ll{eqc}
\eq
with the  coefficients $a,b$ and $c$ depending on the variables
$(x,p_x,y,p_y)$.  There are two solutions $C^2_{1,2}$, which are
related by the change of variables $y\to-y\,,~p_y\to -p_y$, and
they are polynomial expressions in  the new variables.
Notice that the second integral of motion $C$ (\ref{int}),
which is polynomial in $(u,p_u,v,p_v)$ becomes an
algebraic function of $(x,p_x,y,p_y)$.

After substitution of $C(x,p_x,y,p_y)$ into
(\ref{invtr}) we get an inverse transformation.
\begin{prop}
A Hamiltonian
\[H=f(u)p_u^2+g(v)p_v^2+V_1(u)+V_2(v)\]
with the potentials given by (\ref{v12}) transforms to either of
two natural Hamiltonians
\ben
H_1&=&\dfrac12(p_x^2+p_y^2)-\dfrac{\a_2}{2\a_1}(x^4+6x^2y^3+8y^4)\nn\\
\ll{qp1}
&+&\dfrac{\b_2}2 (x^2+4y^2)
+\dfrac{2\b_0}{\a_2^2y^2}+2\a_2\b_1\,,
\en
and
\bq
H_2=\dfrac12(p_x^2+p_y^2)-\dfrac{\a_2}{2\a_1}(x^4+6x^2y^3+y^4)\,,\ll{qp2}
\eq
under the quasi-point transformation (\ref{ptr})
and under the point transformation, respectively.
\end{prop}
For brevity, we put $\a_3=0$ in $H_1$ (\ref{qp1}),
since the introduction of  $\a_3\neq 0$ corresponds to the shift of
the variable $y\to y+\a_3/4\a_2$ \cite{rom95}.  In $H_2$ (\ref{qp2})
we presented only the highest polynomial terms.

Systems with the Hamiltonians $H_1$ and $H_2$ are associated with
restricted flows of some PDE's \cite{brw94,bef95}.  In classical
mechanics these systems have  common separated equations.
Second integrals of motion $C^2$ derived from the equation
(\ref{eqc}) are well known and can be founded in
\cite{rav94,rom95}.

By rescaling constants $a_j,\b_j$ and by taking the limits
$\a_2\to 0$ and $\a_3\to 1$, the Hamiltonians
(\ref{qp1}) and (\ref{qp2}) are transformed to the following
Hamiltonians for the Henon-Heiles system \cite{brw94,rav93}
\ben
H_1&=&\dfrac12(p_x^2+p_y^2)+
\dfrac{2}{\a_1\a_3}y(3x^2+16y^2)
-\b_2(x^2+16y^2)\nn\\
\nn\\
&+&
2(\a_1\b_2-2\b_1)y+
2\b_0+\a_1\b_2\b_1\,,\nn\\
\ll{hhp}\\
H_2&=&\dfrac{p_x^2+p_y^2}2+ay(x^2+2y^2)\,.\nn
\en
A complete account of this limit procedure can be found in
\cite{rom95}.

It is known that the Hamiltonian (\ref{qp1})
has an integrable extensions
\[H=H_{1,2}+\dfrac{m}{x^2}+\dfrac{l}{y^2}
+\dfrac{n}{x^6}+ey\,,\]
where either $e=0$, or $n=m=0$ \cite{rav94}.  We can  include
the terms $\mu x^{-2}$ and $ly^{-2}$ into our proposed scheme.
Let us consider the following infinite-dimensional
representation of $sl(2)$ defined in  Cartan-Weil basis
\[ s_3=\dfrac{xp}2\,,\quad
s_+=\dfrac{x^2}2\,,\quad s_-=-\dfrac{p^2}2\,,
\quad\{p,q\}=1\,.\]
The mapping
\ben
&&s_3\to s_3'=s_3\,,\qquad s_+\to s_+'=s_+\,, \nn\\
\ll{mapp}\\
&&s_-\to s_-'=s_-+fs_+^{-1}=\dfrac{p^2}2+\dfrac{2f}{x^2}\,,
\qquad f\in\bR \nn
\en
is an outer automorphism of the space of
infinite-dimensional representations of $sl(2)$.
For the Henon-Heiles system and for the system with quartic
potential the phase space can be identified completely or partially
with the coadjoint orbits in $sl(2)^*$ as (\ref{mapp}).
Hence, all the presented results
can be carried over on the systems with shifted squared
momenta. The corresponding deformation of the separated
equation has been described in \cite{rav94}.


\section{Quasi-point transformations and SUSY quantum mechanics}
Next, we present the interesting relations of the
quasi-point canonical transformation with the supersymmetrical
quantum mechanics, which represents in a concise algebraic form
the spectral equivalence between different Hamiltonian quantum
systems realized by the Darboux transformation
\cite{aj95}.
Further, for brevity, we fix the value of parameters
$\b_k=0,~\a_2=1\,,~\a_1=2$.

Variables of separation $u,v$ for the Hamiltonian $H_2$ can be
defined as the roots of quadratic equation
\bq
\x^2-y\x+\dfrac{y^2-x^2}4=0\,.\ll{eq1}
\eq
The quasi-point transformations (\ref{invtr}) can be presented
in similar form.
Variables $u,v$ (\ref{invtr}) are roots of the quadratic equation
\bq
\x^2-\dfrac{(q_++q_-)}{x^2}\x+\dfrac{(q_+-q_-)^2}{4x^2}=0\,.
\ll{eq2}
\eq
Here, we introduced functions $q_\pm$ and $f$ with
the following  properties
\bq
\{H, q_\pm\,\}=\pm f\,q_\pm\,,\qquad
C^2=4q_+\,q_-\,.
\ll{susy}
\eq
This algebra (\ref{susy}) is the classical limit of the
two-dimensional quantum SUSY algebra \cite{aj95}.
For  the system with quartic potential functions
$q_\pm$ and $f$ are given by
\ben
&&q_+=\dfrac12p_x^2-\dfrac14x^2(2y^2+x^2)+x(yp_x-\dfrac12p_yx)\,,\nn\\
\nn\\
&&q_-(x,p_x,y,p_y)=q_+(x,p_x,-y,-p_y)\,,\ll{qpm1}\\
\nn\\
&&f=2y\,,\nn
\en
and for the Henon-Heiles system
\ben
&&q_+=\dfrac12 p_x^2+2x^2 y+ix\cdot
\sqrt{\,2(2 p_x y-x p_y)p_x-(8 y^2+x^2)x^2)\,}\,,\nn\\
\nn\\
&&q_-(x,p_x,y,p_y)=q_+^*(x,p_x,y,p_y)\,,\ll{qpm2}\\
\nn\\
&&f=i\dfrac{4p_xy-xp_y}{
\sqrt{\,2(2 p_x y-x p_y)p_x-(8 y^2+x^2)x^2)\,}}\nn
\en
where $i$ is the imaginary unit and $q_+^*$ means the complex
conjugate of the function $q_+$.  For the arbitrary values of the
parameters $\a_k$ and $\b_k$ these functions are derived
by equations (\ref{invtr}) and (\ref{eq2}).

For the system with quartic potentials functions $q_\pm$ and $f$
(\ref{qpm1}) have been introduced in \cite{rgd86}
by considering a linearization of the corresponding Hamiltonian flow
on certain constraint submanifolds.

It would be interesting to apply  equation (\ref{eq2})
and their quantum counterpart for separating variables
in quantum mechanics.

Motivated by relation of classical limit of SUSY quantum mechanics
and quasi-point separation of variables we present the classical SUSY
algebra (\ref{susy}) for another two-dimensional natural Hamiltonian
systems. For the  systems of \cite{wowo84} and for the Holt-like
systems the classical limit of SUSY algebra (\ref{susy}) is
equal to
\ben &&H_{W}=\dfrac{p_x^2+p_y^2}2 +
+\left(\dfrac{a_1^2}2+\a_2\a_3\right)x^{2/3}
-\dfrac9{16}\a_1^2\a_2 x^{-2/3}y-\a_3y\,,\nn\\
\nn\\
&&q_+=
-\dfrac{p_x^2}2+\dfrac9{16}\left(
\a_1^2\a_2 x^{-2/3}y-\dfrac{\a_2}{\a_1}\right)
+\left(\dfrac{\a_1^2}2+\a_2\a_3\right)x^{2/3}
+i\left(\a_1 x^{1/3}p_x+\dfrac{\a_2}{3\a_1}p_y\right)\,,\nn\\
\nn\\
&&f=-\dfrac23 i \a_1 x^{-2/3}\nn
\en
and
\ben
&&H_{H}=\dfrac{p_x^2+p_y^2}2 +y^{-2/3}(\dfrac92 y^2+x^2)\,,\nn\\
\nn\\
&&q_+=p_x^2+2y^{-2/3}x^2+i\cdot
\sqrt{\,2(p_x p_y+6xy^{1/3})^2
-(2x^2y^{-2/3})^2)\,}\,,\nn\\
\nn\\
&&f=\dfrac{4i x^2 p_y}
{3y\sqrt{ 2(y^{2/3}p_xp_y+6yx)^2-4x^4 }}\,,\nn
\en
where
$q_-(x,p_x,y,p_y)=q_+^*(x,p_x,y,p_y)$.

Notice, that the construction of isospectral two-dimensional
Hamiltonians in supersymmetrical quantum mechanics is closely
connected with another problem, namely with a search for the
second integral of motion for the quantum integrable systems \cite{aj95}.
Here we present a new characterisation of this problem.
\begin{prop}
Let us start with the classical SUSY algebra (\ref{susy}) for a
two-dimensional integrable system defined by four functions
$H,f$ and $q^{\pm}$.  If the following equation for $\D h$ can
be solved
\bq
\{\D h,\{q_+,q_-\}\,\}=\{f,q_+\}q_- +\{f,q_-\}q_+\,,\ll{h}
\eq
then the pair of the mutually commuting integrals of motion
\bq
\te{H}=H+\D h,,\qquad
\te{C}=\{q^+,q^-\}\,,\qquad\{\te{H},\te{C}\}=0\,,
\ll{newsys}
\eq
defines a new two-dimensional  integrable system.
\end{prop}
The evolution on a $2n$-dimensional symplectic manifold
is called completely integrable if there exists $n$
functions $I_1,\ldots,I_n$, which are independent integrals
in the involution \[\{I_i(x,p),I_j(x,p)\}=0\,,\qquad i,j=1,\ldots,n\,.\]
The initial integrals of motion $H$ and $C$ are
functionally independent functions and, therefore, the new integrals
$\te{H}$ and $\te{C}$ are independent.
For the proof of involution of integrals
we need the the Jacobi
identity for the Poisson brackets. Assuming first (\ref{susy})
holds we have
\ben
\{H,\te{C}\}&=&\{H,\{q^+,q^-\}\}=\{q^+,\{H,q^-\}\}-
\{q^-,\{H,q^+\}\}=\nn\\
&=&q^-\{f,q^+\}+q^+\{f,q^-\}\,.\nn
\en
Using next (\ref{h}) we get (\ref{newsys})
\[\{H+\D h,\te{C}\}=\{\te{H},\te{C}\}=0\,.\]
So, we have two independent integrals of motion in involution, which
are defined new integrable system.

As an example, for the system with quartic potential
equation (\ref{h}) can be integrated and it yields
\ben
H_1&=&\dfrac{p_x^2+p_y^2}2
-\dfrac14(x^4+6x^2y^2+8y^4)\,,\nn\\
\nn\\
\D h&=&-\dfrac18 x^4\,,\nn\\
\ll{hpar}\\
\te{H}&=&H_1+\D h=
\dfrac{p_x^2+p_y^2}2
-\dfrac18(x^4+12x^2y^2+16y^4)\,,\nn\\
\nn\\
\te{C}&=& \{q_+,q_-\}=16\left(
2(p_yx-yp_x)p_x-(2y^2-x^2)yx^2
\right)\,.\nn
\en
New system with integrals of motion $\te{H},\te{C}$
is separable in the parabolic coordinates, which
are defined as roots of the quadratic equation
$\x^2+2y\x-x^2=0$.

\section{Conclusions}
\setcounter{equation}{0}
In the previous paragraphs we have examined
the six-parameters quasi-point canonical transformation
leading to  separation of variables for
the Henon-Heiles system and a system with quartic potential.
We have proved that this transformation strictly
connected to the rectangular separated equations and that it can not
be generalized. In addition we found some relations of these
variables of separation with the two-dimensional supersymmetric
quantum mechanics.

This work was supported by Swedish Royal Academy of Science grant,
project N1314 for collaboration with Russian Academy of Science,
and RFBR grant.


\end{document}